\documentclass[%
superscriptaddress,
preprint,
 bibnotes,
 amsmath,amssymb,
]{revtex4-1}

\usepackage{graphicx}
\usepackage{dcolumn}
\usepackage{bm}
\usepackage{color}
\usepackage[colorlinks,linkcolor=blue,anchorcolor=blue,citecolor=blue,urlcolor=black]%
{hyperref}

\begin{document}


\title{Charge density waves and phonon-electron coupling in ZrTe$_3$ investigated by Raman spectroscopy and first-principles calculations}

\author{Yuwen~Hu}
\email[These coauthors contributed equally to this work.]{}
\affiliation{International Center for Quantum Materials, School of Physics, Peking University, Beijing 100871, China}
\author{Feipeng~Zheng}
\email[These coauthors contributed equally to this work.]{}
\affiliation{International Center for Quantum Materials, School of Physics, Peking University, Beijing 100871, China}
\author{Xiao~Ren}
\affiliation{International Center for Quantum Materials, School of Physics, Peking University, Beijing 100871, China}
\author{Ji~Feng}
\email[]{jfeng11@pku.edu.cn}
\affiliation{International Center for Quantum Materials, School of Physics, Peking University, Beijing 100871, China}
\affiliation{Collaborative Innovation Center of Quantum Matter, Beijing 100871, China}
\author{Yuan~Li}
\email[]{yuan.li@pku.edu.cn}
\affiliation{International Center for Quantum Materials, School of Physics, Peking University, Beijing 100871, China}
\affiliation{Collaborative Innovation Center of Quantum Matter, Beijing 100871, China}

\begin{abstract}
Charge-density-wave (CDW) order has long been interpreted as arising from a Fermi-surface instability in the parent metallic phase. While phonon-electron coupling has been suggested to influence the formation of CDW order in quasi-two-dimensional (quasi-2D) systems, the presumed dominant importance of Fermi-surface nesting remains largely unquestioned in quasi-1D systems. Here we show that phonon-electron coupling is also important for the CDW formation in a model quasi-1D system ZrTe$_3$. Our joint experimental and computational study reveals that particular lattice vibrational patterns possess exceedingly strong coupling to the conduction electrons, and are directly linked to the lattice distortions associated with the CDW order. The dependence of the coupling matrix elements on electron momentum further dictates the opening of (partial) electronic gaps in the CDW phase. Since lattice distortions and electronic gaps are the defining signatures of CDW order, our result demonstrates that the conventional wisdom based on Fermi-surface geometry needs to be substantially supplemented by phonon-electron coupling even in the simplest quasi-1D case. As prerequisites for the CDW formation, the highly anisotropic electronic structure and strong phonon-electron coupling in ZrTe$_3$ give rise to a distinct Raman scattering effect, namely, measured phonon linewidths depend on the direction of momentum transfer in the scattering process.

\end{abstract}


\maketitle

\section{\label{sec1}Introduction}

The mechanism for charge-density-wave (CDW) order in metals is a long-standing problem in condensed matter physics. The recent discovery of ubiquitous charge-ordering phenomena in cuprate high-temperature superconductors \cite{Wu2011,Ghiringhelli2012,Comin2014,Tabis2014,daSilvaNeto2014,Abbamonte2005,Tranquada1995} has aroused new interest in this problem, since a thorough and generic understanding of the formation of CDW order, even in simple metals, may shed light on several important issues relevant to the enigma of high-temperature superconductivity. In particular, the so-called pseudogap phenomena in underdoped cuprates \cite{Norman2005,Timusk1999}, which precede superconductivity upon cooling, resemble the opening of partial electronic energy gaps on certain parts of Fermi surfaces in CDW metals \cite{Borisenko2009,Brouet2004,Yokoya2005}. While long-range static CDW order appears to compete with superconductivity \cite{Ghiringhelli2012,daSilvaNeto2014,Croft2014}, short-range CDW correlations are present over a substantial range of carrier concentrations \cite{Ghiringhelli2012,LeBoeuf2011} that support the appearance of superconductivity, which points to the intriguing possibility that the two phases might share a common microscopic origin. It is, therefore, important to identify the necessary and sufficient conditions for driving a CDW transition, and to understand how a material's electronic structure can be affected upon approaching such transitions.

CDW order is most commonly found in metals with highly anisotropic, or ``low-dimensional'', electronic structure. Typical examples include quasi-one-dimensional (quasi-1D) K$_2$Pt(CN)$_4$ \cite{Comes1973}, blue bronzes \cite{Pouget1991}, and transition-metal trichalcogenides \cite{Dicarlo1994,Felser1998}, as well as quasi-2D transition-metal dichalcogenides \cite{Tsang1976,Dardel1992} and rare-earth tritellurides \cite{Hu2014}. In the classical explanation first introduced by Peierls \cite{Peierls1955}, CDW transitions are understood as arising from the presence of a pair of nearly parallel Fermi surfaces, known as Fermi-surface nesting (FSN), which causes the dielectric response of conduction electrons to diverge at the nesting wave vector $\mathbf{q}_\mathrm{n}$. Even though a periodic modulation of the electron density can be stabilized only by coupling to the crystal lattice \cite{Gruener1988,Gruener1994}, it is $\mathbf{q}_\mathrm{n}$ that determines the periodicity of the modulation, \textit{i.e.}, in the Peierls picture CDW order is primarily electronically driven. In most quasi-1D and some quasi-2D systems, a good agreement between the CDW wave vector $\mathbf{q}_\mathrm{cdw}$ and $\mathbf{q}_\mathrm{n}$ has indeed been found \cite{Yokoya2005,Ando2005,Laverock2005}, in support of the Peierls picture.

However, similar agreement between $\mathbf{q}_\mathrm{cdw}$ and $\mathbf{q}_\mathrm{n}$ is not supported by highly accurate measurements \cite{Rossnagel2001,Shen2007,Borisenko2009} and calculations \cite{Johannes2006} on the prototypical quasi-2D CDW compound NbSe$_2$. First-principles calculations have cast further doubt on the generic validity of using FSN as the sole criterion for potential Peierls instability \cite{Johannes2008}: on the one hand, FSN leads to a pronounced enhancement in the imaginary part of the electronic susceptibility near $\mathbf{q}_\mathrm{n}$, but the formation of CDW order requires a large real part of the susceptibility which does not solely depend on FSN; on the other hand, the susceptibility enhancement due to FSN is usually weak in real materials, and it is excessively fragile against imperfect nesting, impurity scattering, and thermal-broadening effects. Analysis of FSN provides insights into the role of electronic instability in the formation of CDW order, but it is important to realize that the most predictive signature of potential CDW instability is the presence of soft phonon modes in the initiating phase \cite{Kohn1959}. In this line of thinking, it is only natural to recognize the essential role of phonon-electron coupling in the formation of CDW order, as suggested by several authors \cite{Varma1983,Johannes2008}. This idea has been explicitly tested in quasi-2D materials both theoretically \cite{CastroNeto2001,Johannes2006,Ge2010,Gorkov2012,Eiter2013} and experimentally \cite{Weber2011,Eiter2013,Arguello2015}, and has proved quite successful in explaining the $\mathbf{q}_\mathrm{cdw}$ that deviates from $\mathbf{q}_\mathrm{n}$.

\begin{figure}[!ptbh]
\includegraphics[width=3.3in]{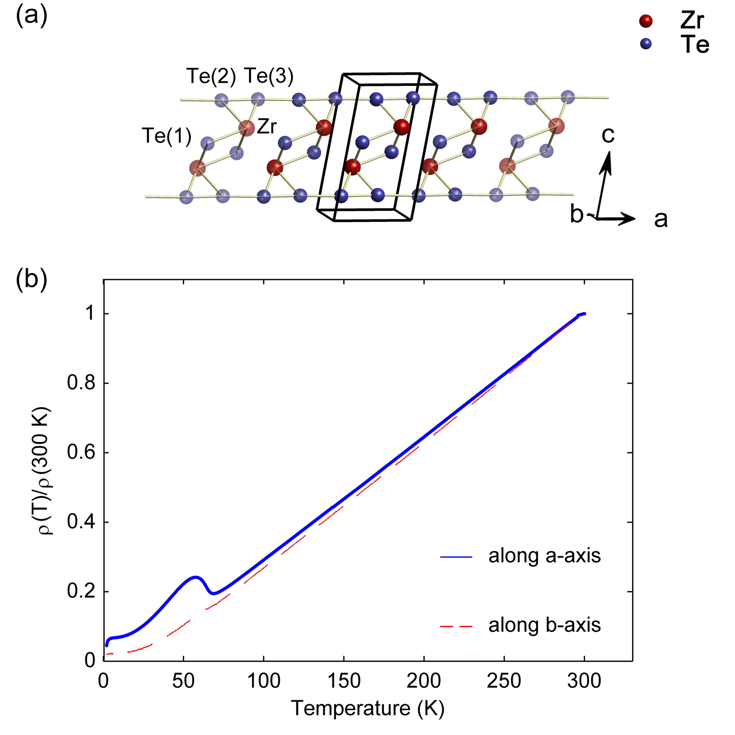}%
\caption{(a) Crystal structure of \textrm{ZrTe$_3$}. (b) Temperature dependence of resistivity measured with currents flowing along the crystallographic $a$ and $b$ axes, normalized at 300 K.}
\label{fig1}
\end{figure}

In quasi-1D materials, the importance of phonon-electron coupling relative to that of FSN has remained largely untested, as the good agreement between $\mathbf{q}_\mathrm{n}$ and $\mathbf{q}_\mathrm{cdw}$ in 1D cases might appear to render such tests unnecessary. One may note, however, that even in a simple quasi-1D system, the Peierls picture lacks such predicting power as to identify lattice distortions that are associated with the CDW order. Moreover, CDW energy gaps are often found on only certain parts of the Fermi surface \cite{Yokoya2005,Ando2005}, even when the remaining parts of the Fermi surface are equally-well nested, which cannot be explained by the Peierls picture. These inadequacies have motivated us to perform a detailed study of phonon-electron coupling in the prototypical quasi-1D CDW compound ZrTe$_3$. ZrTe$_3$ possesses a monoclinic structure with a = 5.89 $\mathrm{\AA}$, b = 3.93 $\mathrm{\AA}$, c = 10.09 $\mathrm{\AA}$, $\alpha = \gamma = 90^{\circ}$, and $\beta = 82.2^{\circ}$ (Fig.~\ref{fig1}(a)). CDW order is stabilized below $T_{\mathrm{cdw}}=63$ K with $\mathbf{q}_\mathrm{cdw} = (0.07 a^{*}, 0, 0.33 c^{*})$ \cite{Seshadri1998}, which is in good agreement with $\mathbf{q}_\mathrm{n}$ that connects quasi-1D Fermi surfaces arising from the $5p$ orbitals of Te(2)/Te(3) atoms that form a chain-like structure along the $a$ axis.

Here we combine variable-temperature Raman scattering measurements and first-principles calculations to investigate the role of phonon-electron coupling in the CDW order in ZrTe$_3$. Despite the material's simple crystal structure, the lattice vibrational properties of ZrTe$_3$ have not been studied using first-principles calculations. A back-to-back comparison of our experimental and computational results yields excellent agreement, which allows us to reliably determine both phonon eigenvectors and phonon-electron coupling matrix elements. We find that only phonons that involve longitudinal movement of Te atoms in the Te(2)-Te(3) chains exhibit large coupling to the conduction electrons. Importantly, we show that the electronic states that are most strongly coupled to the phonons are also those that become gapped out in the CDW phase. These results suggest that certain lattice vibrational patterns play a substantial role in the formation of the CDW order in ZrTe$_3$, and that their interactions with the electrons dictate the opening of partial energy gaps on the Fermi surface. As a joint consequence of the highly anisotropic electronic structure and strong phonon-electron coupling, a distinct Raman scattering effect, namely, the measured phonon linewidths can depend on the direction of momentum transfer in the scattering process, is observed for the first time.

\section{\label{sec2}Methods}
\subsection{\label{sec2.1}Experimental methods}
Single crystals of ZrTe$_3$ were grown by a chemical vapor transport method using iodine as transport gas \cite{Zhu2013}. The crystals were characterized by x-ray powder diffraction and resistivity measurements, which confirmed that our crystals were of single phase and high quality. The x-ray diffraction measurements were performed on a Rigaku MiniFlex diffractometer at room temperature. The resistivity measurement was performed with a standard four-probe method using a Quantum Design PPMS.

Our Raman scattering measurements were performed in a confocal back-scattering geometry on freshly cleaved crystal surfaces using the $\lambda=514$ nm line of an Ar laser for excitation. In order to obtain spectra with light polarizations along various crystallographic axes, crystals were cleaved both parallel and perpendicular to the $ab$ plane. The Raman spectra were analyzed using a Horiba Jobin Yvon LabRAM HR Evolution spectrometer, equipped with 1800 gr/mm gratings, a liquid-nitrogen-cooled CCD detector, and BragGrate notch filters that allow for measurements down to low wave numbers. The temperature of the sample was controlled by a liquid-helium flow cryostat, with the sample kept under better than $5\times10^{-7}$ Torr vacuum at all times. Consistent Raman spectra have been obtained on several different samples, as well as with different excitation-photon wavelengths (633 and 785 nm).

\subsection{\label{sec2.2}Calculational methods}

Density-functional theory calculations were performed using Quantum Espresso \cite{Baroni_link}, within the generalized-gradient approximation parameterized by Perdew, Burke and Ernzerhof \cite{Hedin1971,Perdew1996}, to investigate the electronic structure, BZ-center phonons and phonon-electron coupling in ZrTe$_3$. Norm-conserving pseudopotentials, generated by the method of Goedecker, Hartwigsen, Hutter, and Teter \cite{Hartwigsen1998}, were used to model the interactions between valence electrons and ionic cores of both Zr and Te atoms. The Kohn-Sham valence states were expanded in the plane wave basis set with a kinetic energy truncation at 150 Ry. The Raman-active phonons were calculated at the BZ center using a linear-response approach. The equilibrium crystal structure was determined by a conjugated-gradient relaxation, until the Hellmann-Feynman force on each atom was less than $0.8\times10^{-4}$ eV/$\mathrm{\AA}$ and zero-stress tensor was obtained. A 12$\times$18$\times$8 $\mathbf{k}$-grid centered at the $\Gamma$ point was chosen, in combination with a Gaussian-type broadening of 0.0055 Ry.

The phonon linewidth (defined as full width at half maximum, FWHM, in the Raman spectra), $\gamma_{\bm{q}\nu}$, is twice the imaginary part of the phonon self energy arising from phonon-electron interactions:
\begin{eqnarray}
\gamma_{\mathbf{q}\nu}=\frac{4\pi}{N_k}\sum_{\mathbf{k}mn}|g^{\nu}_{\mathbf{k}n,\mathbf{k+q}m}|^2(f_{\mathbf{k}n}-f_{\mathbf{k+q}m}) \nonumber\\
\delta (\epsilon_{\mathbf{k+q}m}-\epsilon_{\mathbf{k}n} - \omega_{\mathbf{q}\nu}),
\label{Equation1}
\end{eqnarray}
where the summation is over a set of $\mathbf{k}$-points forming a regular grid in the BZ, each with a pair of Kohn-Sham electronic states $m$ and $n$ whose crystal momenta differ by that of the phonon mode $\mathbf{q}$. The phonon-electron coupling matrix elements,
$g^{\nu}_{\mathbf{k}n,\mathbf{k+q}m}= \langle\mathbf{k}n | \delta V/\delta u_{\mathbf{q\nu}} | \mathbf{ k+q}m  \rangle/\ \sqrt{2\omega_{\mathbf{q}\nu}}$,
are obtained from Quantum Espresso \cite{Baroni_link}, where $u_{\mathbf{q}\nu}$ is the atomic displacements of mode $\nu$ at a wave vector $\mathbf{q}$. The Fermi-Dirac distribution, $f_{\mathbf{k}n}$, gives the occupation of the Kohn-Sham state with $\epsilon_{\mathbf{k}n}$. The summation in Eq.(\ref{Equation1}) can be quite singular, and we used the cubic-spline to interpolate the electronic spectra and the squared phonon-electron coupling matrix elements from a ``coarse'' $\mathbf{k}$-grid of $44\times68\times24$ to $660\times1020\times360$, where $\epsilon_{\mathbf{k}n}$ were calculated. Finally, weighted average of $\gamma_{\mathbf{q}\nu}$ from a small region of $\mathbf{q}$ near the BZ center was used to estimate the actual phonon line width measured in Raman experiments  (see Sec.~\ref{sec3.5}).

\section{\label{sec3}Results and discussions}
\subsection{\label{sec3.1}Sample characterization}
Figure~\ref{fig1}(b) displays typical resistivity curves measured on our single crystals in the temperature range between 2 K and 300 K. A clear resistivity anomaly is observed around $T_{\mathrm{CDW}}=63$ K when the electric current flows along the crystallographic $a$ axis. In contrast, a similar resistivity anomaly is very weak or absent in the $b$ axis. Filamentary superconductivity is observed below $T\approx3$ K. These results are consistent with previous reports \cite{Zhu2011,Felser1998}. The small residual resistance at low temperatures demonstrates the high quality of our samples.

\subsection{\label{sec3.2}Identification of Raman-active phonons}
The crystal structure of ZrTe$_3$ belongs to the space group $P12_{1}/m1$, with all atoms occupying Wyckoff $2e$ positions. Group theory analysis shows that there are a total of twelve Raman-active modes in ZrTe$_3$, including eight $A_g$ modes and four $B_g$ modes,  which involve atomic movements in the $ac$ plane and along the $b$ axis, respectively. Only linear photon polarizations are employed in our Raman scattering measurements, and the combination of incident- and scattered-photon polarizations is denoted by two letters that indicate the polarization directions with respect to the crystallographic axes. $A_g$ modes can be observed in $aa$, $bb$, and $cc$ polarizations, whereas $B_g$ modes can be observed in $ab$ and $bc$ polarizations.

\begin{figure}[!ptbh]
\includegraphics[width=3.3in]{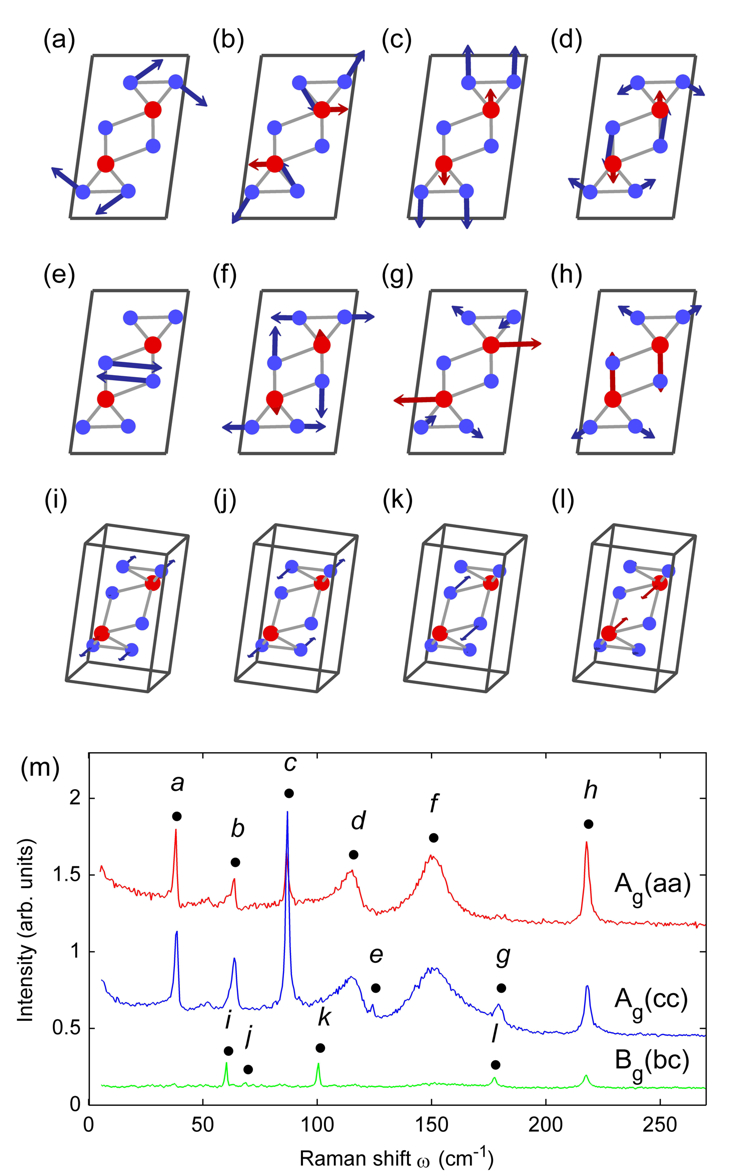}%
\caption{Calculated vibrational patterns of eight $A_g$ (a-h) and four $B_g$ (i-l) phonon modes, from low to high energy. (m) Polarized Raman spectra obtained at 70 K, offset for clarity. Letters indicate the corresponding calculated vibrational patterns.}
\label{fig2}
\end{figure}

Figure~\ref{fig2}(m) displays Raman spectra obtained at $T = 70$ K $>T_\mathrm{cdw}$ in the $aa$, $cc$, and $bc$ polarization geometries. The spectrum obtained in $aa$ geometry is consistent with previous reports \cite{Zwick1980,Gleason2015} and reveals a total of six phonon peaks, but it fails to detect two remaining $A_g$ modes, presumably due to unfavorable Raman-scattering matrix elements. These two modes become clearly visible in the $cc$ spectrum. The spectrum taken in $bc$ geometry reveals all four $B_g$ modes, which are found at different frequencies than those of the peaks in the $aa$ and $cc$ spectra. Cross-leakage signals related to the $A_g$ modes are present in the $bc$ spectrum, but are ignored in our identification of phonons. We note that some of the cross-leakage signals are clearly larger than expected if we only consider the imperfection of our polarizing optical set up ($\sim1\%$), and they are hence likely related to the presence of disorder or defects in our crystals which may remove some of the symmetry selection rules.

The successful observation of all twelve Raman-active phonons in Fig.~\ref{fig2}(m) is crucial for the identification of their eigenvectors. Contrary to the assumption in previous reports \cite{Zwick1980,Gleason2015}, the frequencies of the $A_g$ modes are not well-separated into a low- and a high-frequency group. Therefore the $A_g$ modes cannot be simply attributed to inter- and intra-ZrTe3-prism vibrations. We hence determined the phonon eigenvectors by first-principles calculations, the result of which is displayed in Figures~\ref{fig2}(a)-\ref{fig2}(l), where vibrational patterns are indicated by arrows with lengths proportional to the atomic displacements. Indeed, apart from a distinct difference between the $A_g$ and $B_g$ modes enforced by the structural symmetry, we find no clear regularity in the vibrational patterns. The calculated mode energies are displayed in Table~\ref{table1} along with the measured values at $T = 150$ K. This temperature is chosen such that the phonons are least affected by anharmonic lattice interactions (stronger at higher temperatures) and the formation of CDW correlations (stronger at lower temperatures), which are not accounted for in our calculations. Despite a slight systematic underestimation in the calculated values by no more than 16 cm$^{-1}$, the overall agreement between the experimental and computational results is remarkable, rendering our determination of the phonon eigenvectors highly reliable. A one-to-one correspondence between the calculated eigenvectors and the observed Raman peaks is shown in Fig.~\ref{fig2}.

\begin{table}
\begin{tabular}{ccccccccccc}
\hline
\hline
$A_g$ Phonon & &$a$ &$b$ &$c$ &$d$ &$e$ &$f$ &$g$ &$h$\\
\hline
$\omega_{\mathrm{exp}}$ & &38 &63 &86 &116 &123 &147 &177 &217\\
\hline
$\omega_{\mathrm{cal}}$ & &38 &56 &70 &103 &107 &142 &169 &201\\
\hline
$\gamma_{\mathrm{exp}}$, $\mathbf{q}$ along $\mathbf{a}^*$ & &1.6 &2.5 &1.8 &15 &- &30 &- &3.3\\
\hline
$\gamma_{\mathrm{exp}}$, $\mathbf{q}$ along $\mathbf{c}^*$ & &1.3 &2.4 &2.4 &16 &- &20.5 &- &3.5\\
\hline
$\gamma_{\mathrm{cal}}$, $\mathbf{q}$ along $\mathbf{a}^*$ & &1.5 &0.9 &0.8 &9.8 &0.4 &14.9 &0.3 &3.8\\
\hline
$\gamma_{\mathrm{cal}}$, $\mathbf{q}$ along $\mathbf{c}^*$ & &1.0 &0.9 &0.7 &9.3 &0.4 &10.8 &0.2 &4.5\\
\hline
\hline
$B_g$ Phonon & &$i$ &$j$ &$k$ &$l$\\
\hline
$\omega_{\mathrm{exp}}$ & &60 &67 &100 &176\\
\hline
$\omega_{\mathrm{cal}}$ & &56 &65 &103 &170\\
\hline
\end{tabular}
\caption{Experimental and calculated phonon frequencies and linewidths at 150 K. All energy values are in units of cm$^{-1}$. The experimental determination of linewidth for modes $e$ and $g$ has large uncertainties because of their weak signal.}
\label{table1}
\end{table}

\subsection{\label{sec3.3}Phonon linewidths}
\begin{figure}[!ptbh]
\includegraphics[width=3.3in]{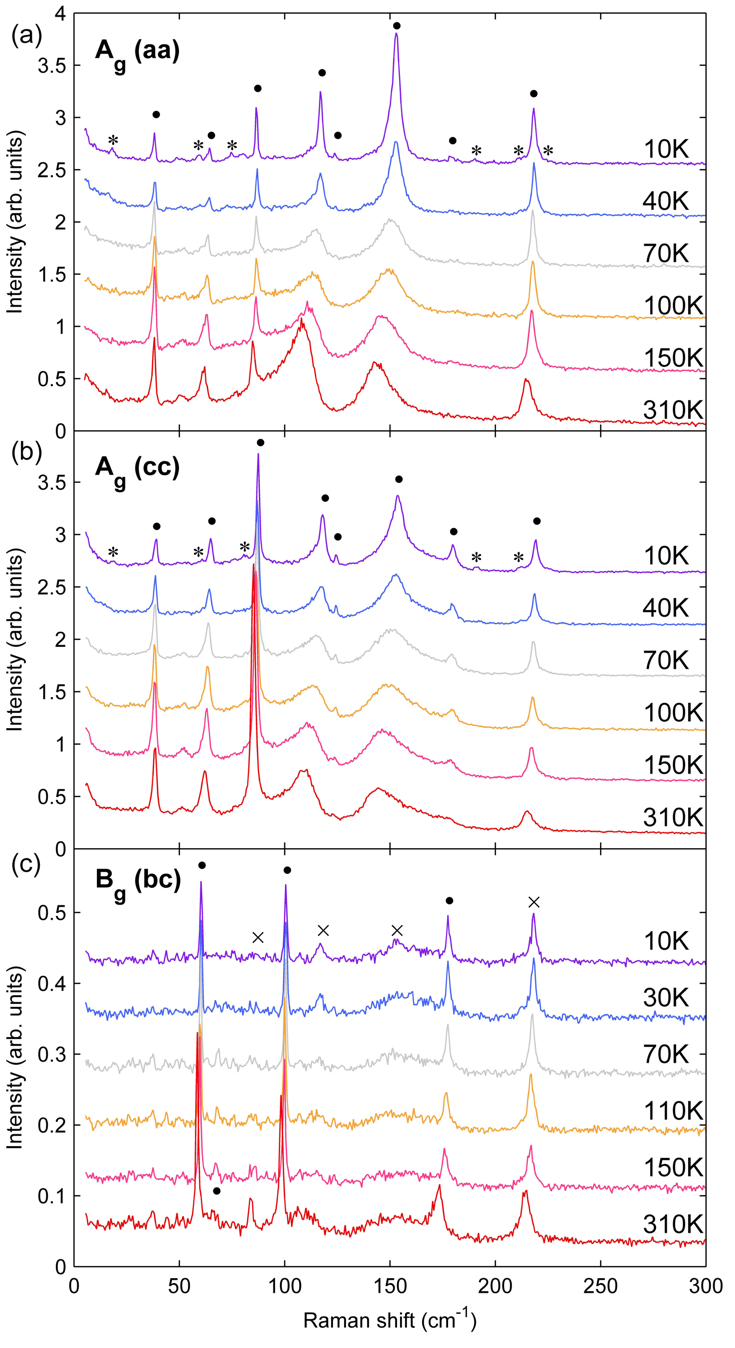}%
\caption{Raman spectra obtained with different photon polarizations at selected temperatures, offset for clarity. ``$\bullet$'' indicates phonon peaks that can be indexed by our calculational results shown in Fig.~\ref{fig2}, ``$\ast$'' indicates additional peaks that appear below $T_\mathrm{cdw}$, and ``$\times$'' indicates signals related to cross leakage between different polarizations.}
\label{fig3}
\end{figure}

\begin{figure}[!ptbh]
\includegraphics[width=3.3in]{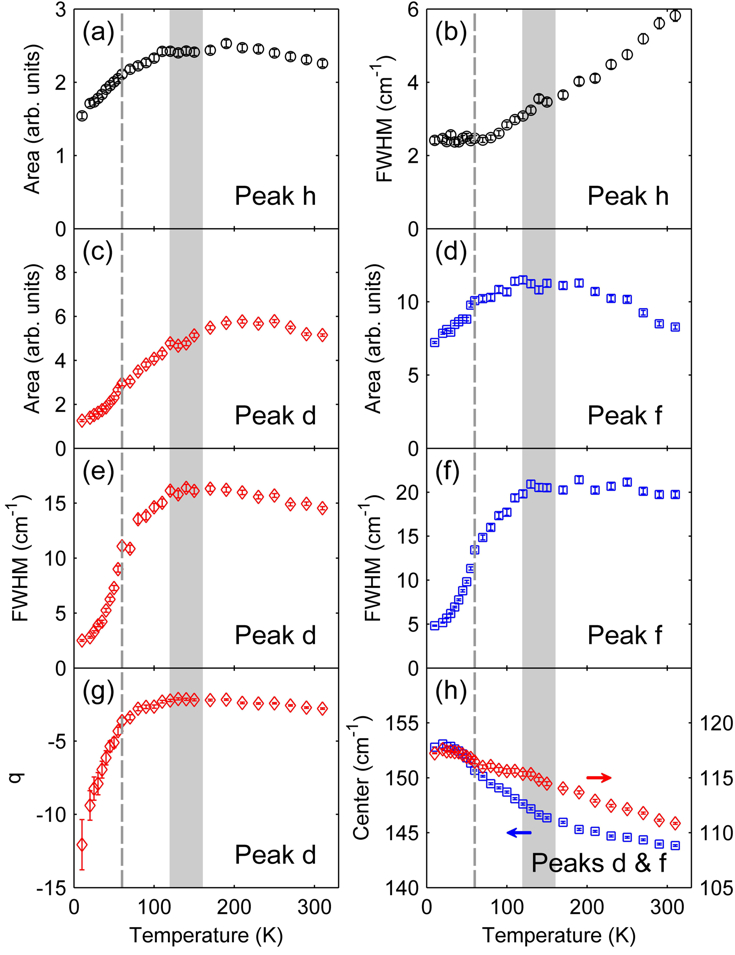}%
\caption{Temperature dependence of fitting parameters of selected phonons measured with $aa$ polarizations. The peak indices ($d$, $f$, and $h$) are after those in Fig.~\ref{fig2}. Vertical dashed line indicate $T_\mathrm{cdw}$. Shaded area indicates a higher characteristic temperature $T^*$ which can only be determined with a large uncertainty. In (a), (c), and (d), the peak areas are displayed after divided by the Bose factor.}
\label{fig4}
\end{figure}

An important observation in Fig.~\ref{fig2}(m) is that the $d$ and $f$ phonon modes exhibit much larger linewidths than the rest. Moreover, the peak of mode $d$ is distinctly asymmetric and can be better described by a Fano than a Lorentzian function. It therefore seems likely that these two phonons are strongly coupled to a continuum of excitations, such as electronic excitations. To test this possibility we have performed variable-temperature Raman measurements (Fig.~\ref{fig3}). By fitting the data to Lorentzian (mode $f$) and Fano (mode $d$) functions, we extract the evolution of peak parameters as a function of temperature (Fig.~\ref{fig4}). The fitted parameters of mode $h$, which does not exhibit an unusual linewidth, are also shown in Fig.~\ref{fig4} for comparison.

As temperature decreases, the linewidths (FWHM) of modes $d$ and $f$ are found to decrease substantially (Figs.~\ref{fig4}(e) and \ref{fig4}(f)): the decrease starts already from above $T_\mathrm{cdw}$ below a characteristic temperature $T^* \approx 140$ K, and is particularly strong near $T = T_\mathrm{cdw}$. Meanwhile, as shown in Fig.~\ref{fig4}(g) the asymmetry parameter $q$ in the Fano function for mode $d$ increases sharply in its magnitude as temperature is lowered below $T_\mathrm{cdw}$, which indicates that the mode rapidly recovers towards a Lorentzian lineshape as the CDW order develops. Importantly, we find that both modes $d$ and $f$ involve a longitudinal deformation of the Te(2)-Te(3) chains in their vibrational eigenvectors (Fig.~\ref{fig2}), which will be discussed in more details later. For mode $h$, which does not exhibit a broad linewidth, the fitted FWHM evolves smoothly through $T^*$ and $T_\mathrm{cdw}$ (Fig.~\ref{fig4}(b)). Also the peak positions of modes $d$ and $f$ do not show clear anomalies near these temperatures (Fig.~\ref{fig4}(h)).

We attribute the higher characteristic temperature $T^*$ to the onset of CDW correlations in ZrTe$_3$. A higher onset temperature of short-range correlations than the actual phase transition temperature can be generally expected in low-dimensional systems due to fluctuations. The decrease of phonon linewidths below $T^*$ thus suggests that the electronic states that are coupled to the phonons start to be removed from the Fermi surface by the incipient short-range CDW correlations. This is in qualitative agreement with ARPES results \cite{Yokoya2005}, in which $T^*$ is estimated to be above 200 K. In addition to the phonon linewidths, all the phonon intensities (integrated area under the peaks) show a clear departure from the high-temperature behavior below $T^*$ (Figs.~\ref{fig4}(a), \ref{fig4}(c), and \ref{fig4}(d)), which suggests that the local structure of the crystal lattice starts to deform along with the development of CDW correlations.

\subsection{\label{sec3.4}Electronic Raman signal}
\begin{figure}[ptbh]
\includegraphics[width=3.3in]{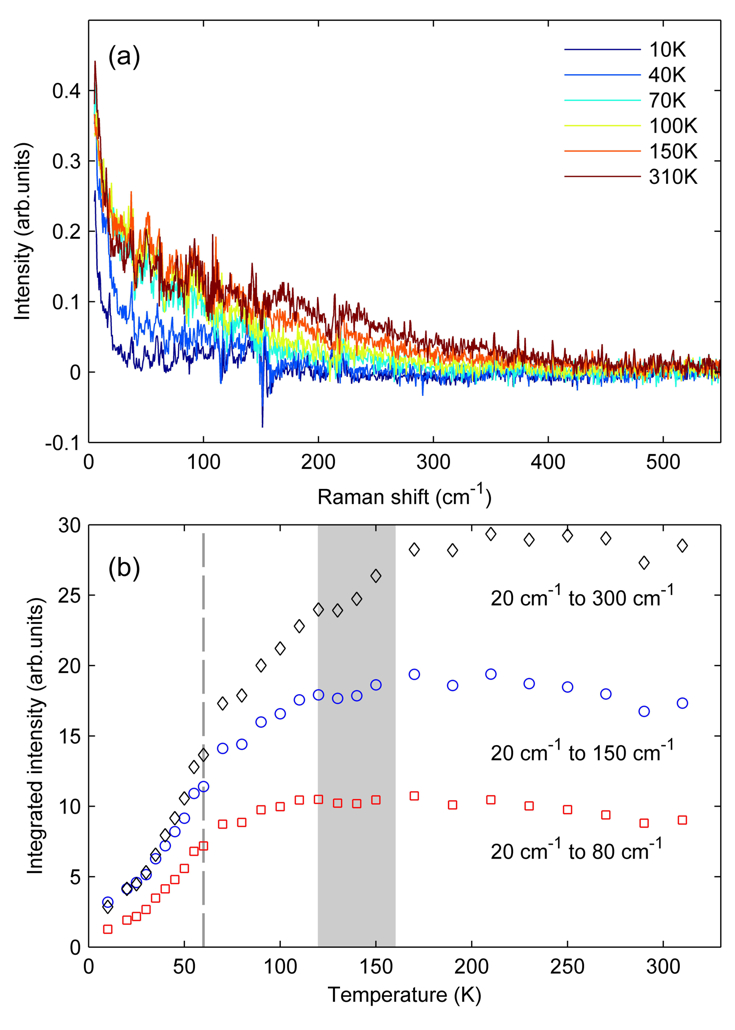}%
\caption{(Color online) (a) $aa$-polarized Raman data obtained at selected temerpatures after subtracting all phonon peaks from the spectra. (b) Integrated intensities in (a) over different energy windows. Dashed line and shaded area indicate $T_\mathrm{cdw}$ and $T^*$, respectively, same as in Fig.~\ref{fig4}.}
\label{fig5}
\end{figure}

To further examine the relationship between the opening of electronic energy gaps due to CDW correlations and the change in phonon-electron coupling, we first note that, especially in the $aa$-polarized Raman spectra in Fig.~\ref{fig3}(a), the background intensity exhibits an anomalous decrease with decreasing temperature. This has also been pointed out in a recent report \cite{Gleason2015}. The highly accurate data in Fig.~\ref{fig3} allow us to fit all the phonon peaks and systematically remove them from the Raman spectra, a procedure that most clearly reveals the underlying electronic signals. Upon doing so, we find that there are further structures in the data, as shown in Fig.~\ref{fig5}(a). Below $\sim100$ cm$^{-1}$, the change of the electronic signal mainly occurs below $T_\mathrm{cdw}$, whereas the change between $\sim100$ and 300 cm$^{-1}$ occurs over a much wider temperature range, up to $T^*$. This can be seen from integration of the electronic signal over different energy ranges (Fig.~\ref{fig5}(b)).

\begin{figure}[!ptbh]
\includegraphics[width=3.3in]{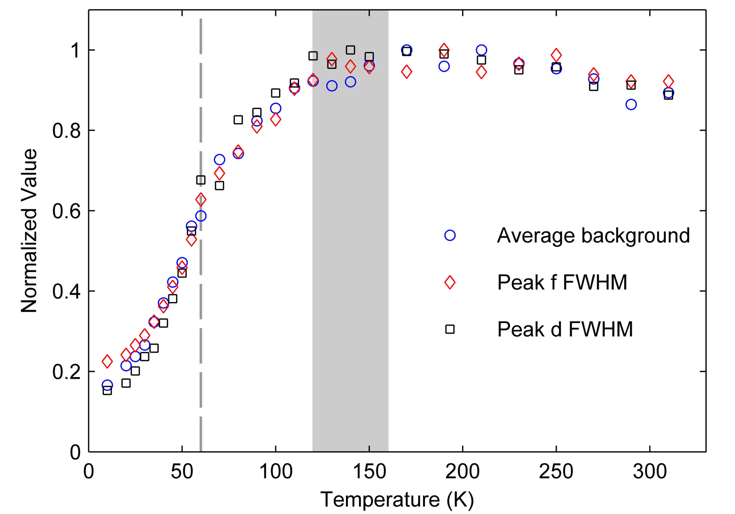}%
\caption{Temperature dependence of the $aa$-polarized electronic signal integrated between 20 and 150 cm$^{-1}$, plotted together with that of the FWHMs of peaks $d$ and $f$ (Fig.~\ref{fig3}(a)). Data are normalized at high temperatures.}
\label{fig6}
\end{figure}

\begin{figure}[!ptbh]
\includegraphics[width=3.3in]{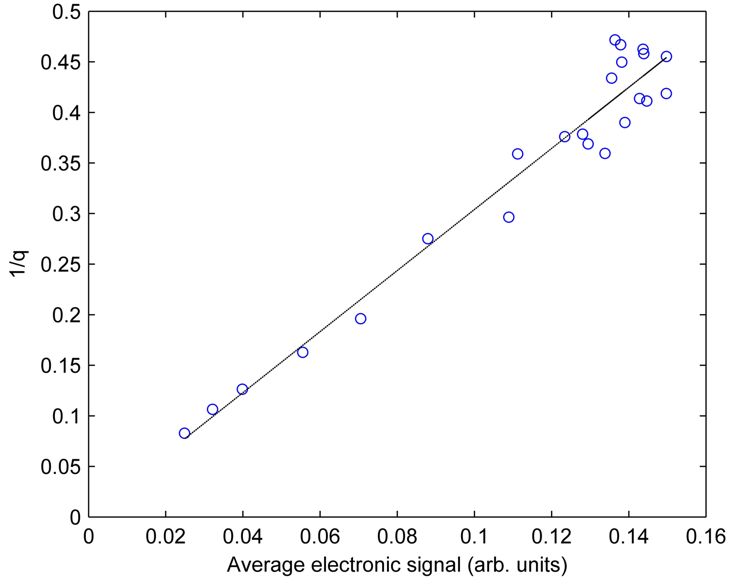}%
\caption{The inverse of the Fano asymmetry parameter for phonon $d$ versus the $aa$-polarized electronic Raman signal averaged from 20 to 150 cm$^{-1}$, with temperature being the implicit parameter. Solid line is a linear fit to the data.}
\label{fig7}
\end{figure}

Being able to extract the electronic signal, we have attempted to correlate it with the change of phonon-electron coupling versus temperature. We first notice that the electronic signal (Fig.~\ref{fig5}(b)) exhibits a similar $T$-dependence as the linewidths of phonons $d$ and $f$ (Figs.~\ref{fig6}(e) and \ref{fig6}(f)). Upon changing the energy-integration window (Fig.~\ref{fig5}(b)), the degree of similarity varies, and the best agreement is found when we integrate the electronic signal up to 150 cm$^{-1}$. This ``optimal'' upper energy limit is in fact commensurate with the phonon energy scale. It implies that the phonons are damped by electronic excitations that are of lower energies than the phonon energies themselves. In addition to the linewidths, we find that the inverse of the Fano asymmetry parameter $q$ for phonon $d$, which describes the peak's departure from a Lorentzian lineshape due to coupling to an excitation continuum \cite{Fano1961}, is roughly proportional to the electronic signal's amplitude over nearly a decade of change (Fig.~\ref{fig7}). While this follows from the fact that the $T$ dependence of $q$ is not very different from that of the peak FWHM (Figs.~\ref{fig4}(e) and \ref{fig4}(g)), it is reassuring to see that $1/q$ extrapolates to zero in the limit of no electronic excitations -- the peak would fully recover to a Lorentzian in the complete absence of phonon-electron coupling.

\subsection{\label{sec3.5} Momentum-resolved phonon-electron coupling matrix elements}

\begin{figure*}[!ptbh]
\includegraphics[width=17cm]{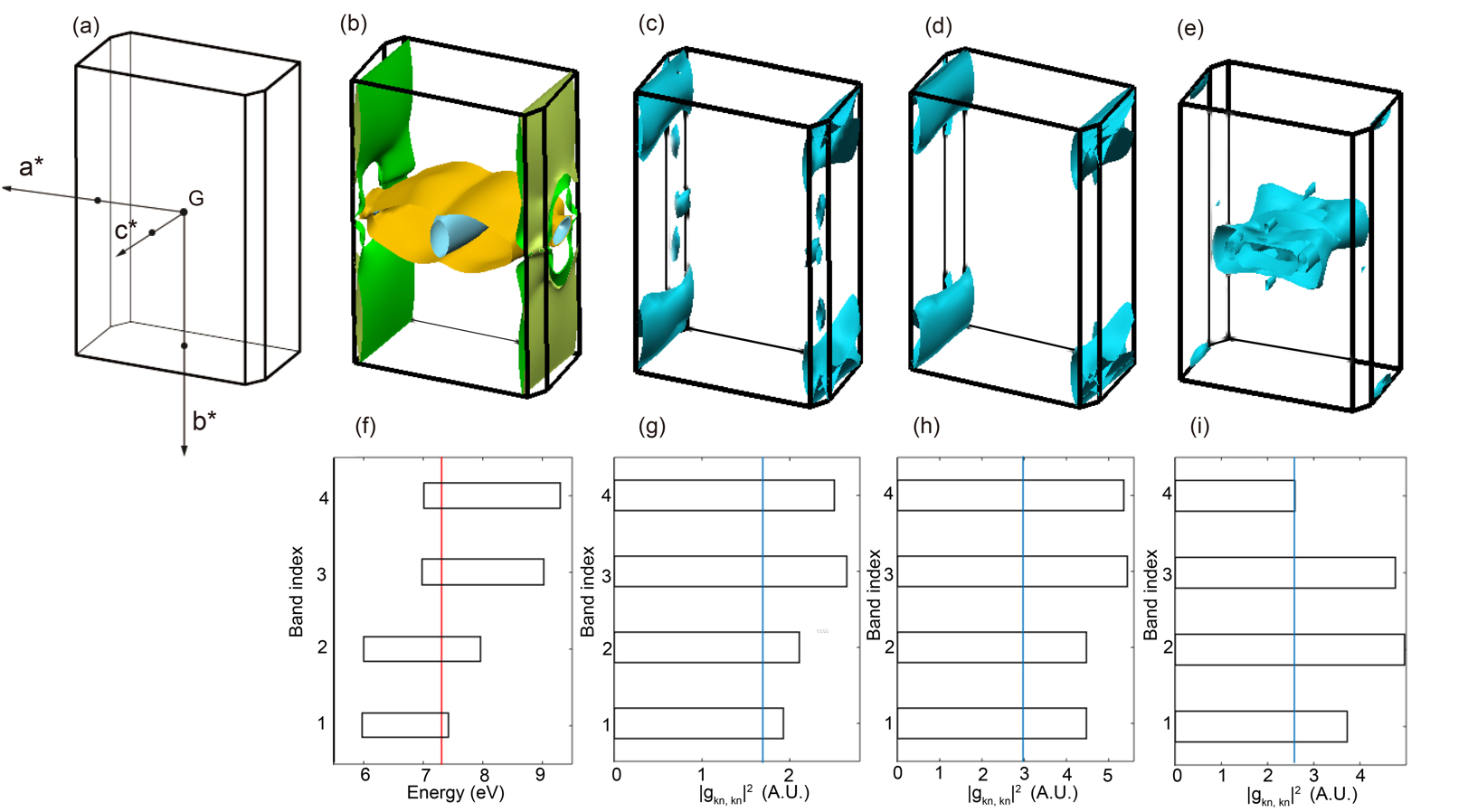}%
\caption{(a) BZ of ZrTe$_3$. (b) Calculated Fermi surfaces. (f) is the corresponding energy locations of the bands across Fermi level (vertical line), marked by indices 1, 2, 3, and 4. (c)-(e) Isosurfaces of $|g^{\nu}_{\bm{k}n,\bm{k}n}|^2$ for phonon modes $d$, $f$, and $h$ in the BZ, with contributions from different bands relative to the chosen isovalues (vertical line) displayed in (g)-(i), respectively. The values of $|g^{\nu}_{\bm{k}n,\bm{k}n}|^2$ are magnified to similar amplitudes for easier visualization. The angle of view in (b)-(e) is similar to that in (a).}
\label{fig8}
\end{figure*}

In order to reach a concrete and microscopic understanding of the experimental results presented in the previous Sections, we have computed the phonon-electron coupling matrix elements for phonons near the BZ center, as functions of the electronic band index and momentum $\mathbf{k}$. Figure~\ref{fig8}(b) displays our computed Fermi surfaces of ZrTe$_3$, which consist of two main sectors as has been previously reported \cite{Yokoya2005,Felser1998}: A quasi-1D sector forms nearly parallel slabs running perpendicular to $\mathbf{a}^*$, which would indeed favor nesting with $\mathbf{q}_\mathrm{n} \approx \mathbf{q}_\mathrm{cdw} = (0.07, 0, 0.33)$. The other sector is located primarily in the vicinity of the $a^*-c^*$ plane. Four electronic bands contribute to the Fermi surfaces. They are marked with indices 1, 2, 3, and 4 in Fig.~\ref{fig8}(f), where their energy ranges are shown relative to the Fermi level. The quasi-1D sector of the Fermi surface located near the BZ boundary is derived from bands 3 and 4, which mainly arise from the $5p$ orbitals of the Te(2)/Te(3) atoms that form a chain-like structure along the $a$ axis (Fig.~\ref{fig1}(a)).

To reproduce the experimentally observed phonon linewidths, it is critical to realize that Raman scattering probes phonons in the small- but non-zero-$\mathbf{q}$ regime. In this regime, while the phonon-electron coupling matrix elements are well approximated by the $q=0$ values, the summation of $\gamma$ in Eq.~(\ref{Equation1}) differs dramatically from the $q=0$ limit owing to intraband transitions. To account for this effect, we have weight-averaged $\gamma$ evaluated at a set of small $\mathbf{q}$ near $q=0$, with a maximum of $|q| = 6\times10^{-3}a_{0}^{-1}$ ($a_{0} =0.5292 \;\mathrm{\AA}$ is the Bohr radius) which corresponds to the back-scattering of 514 nm photons. An index of refraction of 4.6 is used for calculating the momentum carried by 514 nm photons propagating inside ZrTe$_3$ \cite{Herr1986}, and it is assumed that the matrix elements in this small region of $\mathbf{q}$ are the same as those at $q=0$. For simplicity, the weight we used in the $\mathbf{q}$-averaging is just $\gamma$ itself. This way of weighing can be justified by the following argument. Both phonon-electron coupling and Raman process observe energy and momentum conservation, among other conservation laws. The computed phonon-electron matrix elements and hence $\gamma_{\mathbf{q}}$ naturally conforms to this constraint, and is therefore used as a first approximation to the likelihood of the particular electronic transitions involved in the Raman process.

The calculated phonon linewidths are listed in Table~\ref{table1} along with experimental values. A semi-quantitative agreement is found between our experimental and computational results, and in particular we are able to obtain the three broadest peaks correctly in the computation, a result that does not depend on the detailed choice of the $\mathbf{q}$ region or the weighing function. This renders our determination of the phonon-electron coupling matrix elements highly reliable.

Indeed, we find that the computed phonon linewidths are dominated by intraband transitions, especially within bands 3 and 4. An electronic wave vector $\mathbf{k}$ that makes large contributions to the linewidth should satisfy two conditions: (1) the matrix element $|g^{\nu}_{\bm{k}n,\bm{k}n}|^2$ is large, and (2) the electronic energy $\epsilon_{\bm{k}n}$ is sufficiently close to the Fermi level. To delineate the contributions to phonon linewidths from individual bands and different sectors of the Fermi surface, isosurfaces of intraband-transition contributions to the matrix elements for phonons $d$, $f$, and $h$ are visualized in Figs.~\ref{fig8}(c)-\ref{fig8}(e), respectively. Figures~\ref{fig8}(g)-\ref{fig8}(i) display the corresponding value ranges of $|g^{\nu}_{\bm{k}n,\bm{k}n}|^2$ arising from bands $n=$ 1--4, relative to the isovalue for making the isosurfaces. The isosurfaces undergo continuous reduction when the isovalue is increased, and in particular for the phonons $d$ (Fig.~\ref{fig8}(c)) and $f$ (Fig.~\ref{fig8}(d)), they eventually disappear near the corners of the BZ. This shows, strikingly, that the phonons $d$ and $f$ are most significantly coupled to the electronic states near the boundary of the quasi-1D Fermi surfaces. It is not a coincidence that the CDW electronic gap is opened precisely in this region of the BZ \cite{Yokoya2005}, and we believe that it is this region of the BZ that makes the greatest contribution to the overall phonon-electron coupling, including to that of the acoustic phonon whose softening to zero frequency triggers the CDW transition (see Sec.~\ref{sec3.6}). The correspondence between the $\mathbf{k}$ regions shown in Figs.~\ref{fig8}(c)-\ref{fig8}(d) and the gap-opening regions reported in Ref.~\onlinecite{Yokoya2005} suggests that the phonon-electron coupling matrix elements dictates the opening of electronic gaps on the Fermi surface.

A remarkable commonality between phonons $d$ and $f$ is the involvement of longitudinal deformations of the Te(2)-Te(3) chains in their vibrational eigenvectors (Fig.~\ref{fig2}). In contrast, phonons that do not involve similar atomic displacements all have narrow linewidths. This can be understood at an intuitive level, since it is the Te(2)-Te(3) chain that makes a major contribution to the electronic states near the Fermi level. The only exception is mode $h$, which also involves a deformation of the Te(2)-Te(3) chain but does not exhibit a very large linewidth in the Raman spectra. This is explained by the fact that the main contributions to $|g^{\nu}_{\bm{k}n,\bm{k}n}|^2$ are from other regions of the BZ (Fig.~\ref{fig8}(e)), and a considerable portion of those are not on the Fermi surface (Fig.~\ref{fig8}(b)). It is hence no surprise that the linewidth of mode $h$ does not show any pronounced anomaly near $T_\mathrm{cdw}$ or $T^*$ (Fig.~\ref{fig4}(b)); in contrast, a dramatic decrease of the linewidths and a recovery of a Lorentzian lineshape are found for mode $d$ and $f$ below $T_\mathrm{cdw}$ (Fig.~\ref{fig3}).

A detailed comparison of the $aa$- and $cc$-polarization Raman spectra in Fig.~\ref{fig2}(m) and Fig.~\ref{fig3} suggests that the measured linewidth of phonon $f$ is different between the two configurations. In our backscattering experimental geometry, these two configurations correspond to the use of incident photons propagating along the $\mathbf{c}^*$ and $\mathbf{a}^*$ directions, respectively. This difference in linewidth is rather unexpected, and we are not aware of any previous report of similar effects. Our result suggests that the phonon linewidth sensitively depends on $\mathbf{q}$ very close to the BZ center. This dependence can actually be qualitatively reproduced in our computation: by choosing spherical $\mathbf{q}$ regions that are of the same size but centered around different positions relative to the $\Gamma$ point for computing the weight-average of $\gamma$, we find a $50\%$ larger linewidth for phonon $f$ when the center resides along $\mathbf{a}^*$ (Table~\ref{table1}). A pictorial understanding of this result follows from the BZ regions visualized in Fig.~\ref{fig8}(d). The Fermi velocity $\mathbf{v}_\mathrm{F}$ in these regions are primarily along $\mathbf{a}^*$. According to Eq.~(\ref{Equation1}), phonons with $\mathbf{q}$ vectors that satisfy $\mathbf{q}\cdot\mathbf{v}_\mathrm{F} \approx \omega_{\mathbf{q}\nu}$ will have the largest $\gamma$ due to intraband transitions. On the $f$ phonon branch, we find that the $cc$-polarization measurement samples more of such phonons than the $aa$-polarization measurement, and hence observes a larger averaged linewidth.

Despite the apparent similarity in the matrix elements of phonons $d$ and $f$ (Fig.~\ref{fig8}(c)-\ref{fig8}(d)), it turns out that the sampling for phonon $d$ does not have a strong dependence on the center of the $\mathbf{q}$ region considered in our computation. While this result nicely corresponds to our experimental observations (Table~\ref{table1}), it does depend on computational details including the precise band structure near the Fermi level, the shape and size of the sampling $\mathbf{q}$ region, and the weighing function, all of which are not known \textit{a priori}. Therefore the agreement should be taken only at a semi-quantitative level.

According to the above interpretation, the observation of different phonon linewidths with different incident-light directions requires the material system to have specific characteristics. First, the phonon linewidth must be large, and it must arise primarily from phonon-electron coupling that involves electronic intraband transitions (\textit{i.e.}, not from defects or anharmonic lattice interactions). Second, the main contributions to the coupling must come from a highly anisotropic part of the Fermi surface, such that the distribution of pertinent $\mathbf{v}_\mathrm{F}$ has a preferential direction. Failure to conform with the second requirement explains why phonon $h$ does not exhibit a similar behavior as phonon $f$ -- the coupling matrix elements are large on the isotropic sectors of the Fermi surface (Figs.~\ref{fig8}(b) and \ref{fig8}(e)). Finally, we note that the above two requirements are in perfect accordance with the characteristics of a ``good'' CDW system, so it is not a coincidence that we can observe the effect in ZrTe$_3$, which has arguably the simplest crystal structure (and hence phonon spectrum) among all quasi-1D CDW materials known to date.

\subsection{\label{sec3.6}The CDW amplitude excitation}
So far we have only discussed $q=0$ phonons which cannot be directly responsible for the CDW transition. In a recent x-ray scattering measurement of ZrTe$_3$, a pronounced Kohn anomaly was found on the mostly transverse acoustic phonon branch with polarization along the $\mathbf{a}^*$ direction \cite{Hoesch2009}. The freezing of this mode to zero frequency was suggested to be the triggering factor for the formation of the CDW order. To further confirm this interpretation, we have searched for so-called CDW amplitude excitation in the $A_g$ Raman spectra \cite{Cummins1990}. Amplitude excitations can be generally expected in materials that undergoes incommensurate structural transitions including CDW and other forms of charge-ordering phase transitions \cite{Scott1974,Cummins1990,Gruener1994,Du2014}. Such transitions can be generically understood as triggered by an anomalous phonon that freezes to zero frequency at the transition temperature. Deeply in the structurally distorted phase, because the structural instability becomes fully relaxed and/or the electronic states that exert damping on the anomalous phonon become mostly gapped, the frequency of the amplitude excitation is usually found to be close to that of the original (unrenormalized) phonon. In this regard, measurement of the amplitude-excitation frequency can be used to identify the key phonon. To the best of our knowledge, CDW amplitude excitation has not been reported for ZrTe$_3$.

\begin{figure}[ptbh]
\includegraphics[width=3.3in]{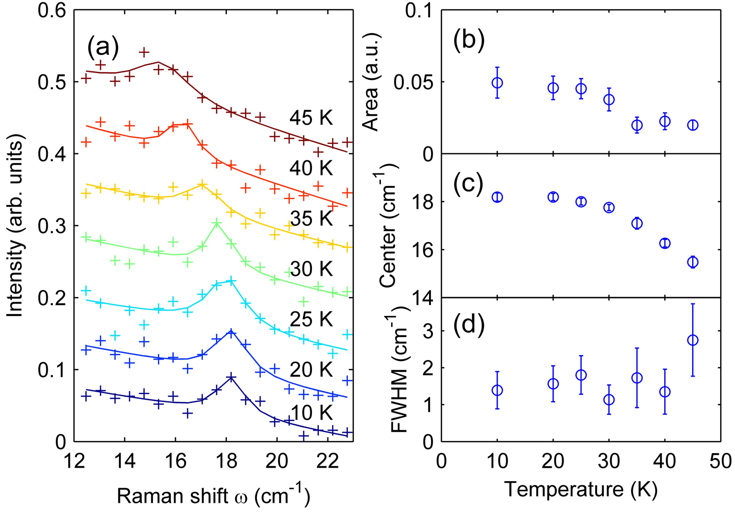}%
\caption{(a) Raman spectra obtained with $aa$-polarizations at low temperatures, offset for clarity. (b)-(d) Temperature dependence of fitting parameters in (a). The peak areas in (b) are displayed after dividing by the Bose factor.}
\label{fig9}
\end{figure}

Indeed, a close inspection of the low energy part of the Raman spectra in Figs.~\ref{fig3}(a)-\ref{fig3}(b) suggests the development of a new peak below 20 cm$^{-1}$ at $T=10$ K. This feature has been measured with very high counting statistics at low temperatures, as shown in Fig.~\ref{fig9}(a), and above $T=45$ K we can no longer resolve it above the diffused-scattering background. By fitting the peak to a Lorentzian profile, we have obtained its key parameters as functions of temperature (Fig.~\ref{fig9}(b)-\ref{fig9}(d)). In particular, its energy softens by over $10\%$ upon heating from 10 K to 45 K. While such an amount of softening is considerably smaller than those observed in quasi-2D \cite{Tsang1976} and 3D materials \cite{Du2014}, it is not uncommon among quasi-1D systems \cite{Sagar2008}, in which the effect of fluctuations is expected to strongly reduce the actual transition temperature from its mean-field value. Therefore we attribute this feature to the CDW amplitude excitation. A similar feature can be identified also in the spectra obtained with $cc$-polarizations (Fig.~\ref{fig3}(b)), but it becomes very difficult to resolve above $T=10$ K because of its weak intensity. Importantly, the amplitude-mode frequency (18 cm$^{-1}$) at 10 K amounts to about 2.2 meV, not far from the unrenormalized energy ($\approx 2.5$ meV) of the Kohn anomaly at high temperatures \cite{Hoesch2009}. This further confirms that, indeed, it is the acoustic phonon that involves the deformation of the Te(2)-Te(3) chains along the $a$ axis that triggers the CDW transition at 63 K.

\section{\label{sec4}Concluding remarks}
In summary, we have performed polarized Raman measurements on ZrTe$_3$, the results of which are compared back-to-back with our first-principles calculations. The latter has allowed us to unequivocally determine the phonon eigenvectors as well as the phonon-electron coupling matrix elements, which have remained hitherto unknown for this model quasi-1D CDW compound. Our result demonstrates that particular lattice vibrational patterns, namely, the longitudinal deformations of the Te(2)-Te(3) chains, exhibit strong interactions with the conduction electrons. Such interactions might play an essential role in the formation of CDW order, even in our simple quasi-1D case where FSN has long been considered to be of dominant importance.

We emphasize that our calculations are for $q\approx0$ phonons only. Our results about the phonon-mode- and electron-$\mathbf{k}$-dependence of the phonon-electron coupling matrix elements do not rely on the FSN geometry, but such dependences should be important to consider for phonons with $\mathbf{q}\approx\mathbf{q}_\mathrm{cdw}$ as well, as they will cast a strong influence on the formation of CDW order. In particular, we believe that it is the matrix elements that select out the specific phonon that first freezes to zero frequency, and hence determine the lattice distortions that are associated with the CDW order. Moreover, it is highly likely that the opening of (partial) electronic gaps on the Fermi surface in the CDW phase is dictated by the $\mathbf{k}$ dependence of the matrix element for the selected phonons, especially when the quality of nesting does not vary substantially across a larger portion of the Fermi surface. Calculations for phonons with $\mathbf{q}\approx\mathbf{q}_\mathrm{cdw}$ warrant further investigations and appear necessary for a quantitative understanding of CDW order even in quasi-1D materials.

As a joint consequence of the highly anisotropic electronic structure and strong phonon-electron coupling in ZrTe$_3$, we have identified a distinct Raman scattering phenomenon, where the measured phonon linewidths can be pronouncedly dependent on the direction of the incident light, contrary to the usual understanding that Raman scattering is a $q=0$ probe. Similar effects can be expected to exist in other low-dimensional CDW materials, where the pertinent requirements are most likely to be conformed simultaneously.

\begin{acknowledgments}

We wish to thank N. L. Wang and F. Wang for stimulating discussions, and L. C. Wang, C. L. Zhang, and D. W. Wang for their assistance in the synthesis and characterization of ZrTe$_3$ samples. The computational work was performed on TianHe-1(A) at the National Supercomputer Center in Tianjin. This work is supported by the NSF of China (No. 11374024 and  11174009) and the NBRP of China (No. 2013CB921903,  2011CBA00109 and 2013CB921900).

\end{acknowledgments}

\bibliography{ZrTe3Ref}

\end{document}